\documentclass[conference]{IEEEtran}
\IEEEoverridecommandlockouts
\usepackage{cite}
\usepackage{amsmath,amssymb,amsfonts}
\usepackage{algorithmic}
\usepackage{graphicx}
\usepackage{textcomp}
\usepackage{xcolor}
\usepackage[caption=false]{subfig}
\usepackage[utf8]{inputenc}
\usepackage{booktabs}
\usepackage{url}
\usepackage{pifont}
\usepackage{hyperref}
\usepackage{tikz}

\newcommand{\xmark}{\ding{55}}%

\hypersetup{
    colorlinks=true,
    linkcolor=black,
    citecolor=black,    
    urlcolor=black,
}

\def\BibTeX{{\rm B\kern-.05em{\sc i\kern-.025em b}\kern-.08em
    T\kern-.1667em\lower.7ex\hbox{E}\kern-.125emX}}
    
\newcommand\copyrighttext{%
  \footnotesize \textcopyright 2020 IEEE. Personal use of this material is permitted. Permission from IEEE must be obtained for all other uses, in any current or future media, including reprinting/republishing this material for advertising or promotional purposes, creating new collective works, for resale or redistribution to servers or lists, or reuse of any copyrighted component of this work in other works. Published in: IEEE Transactions on Neural Networks and Learning Systems; DOI: \url{10.1109/TNNLS.2020.2985720}}
\newcommand\copyrightnotice{%
\begin{tikzpicture}[remember picture,overlay]
\node[anchor=south,yshift=10pt] at (current page.south) {\fbox{\parbox{\dimexpr\textwidth-\fboxsep-\fboxrule\relax}{\copyrighttext}}};
\end{tikzpicture}%
}        
      
\begin{document}

\title{LSTM-MSNet: Leveraging Forecasts on Sets of Related Time Series with Multiple Seasonal Patterns
}
\author{
\IEEEauthorblockN{Kasun Bandara,
Christoph Bergmeir,
Hansika Hewamalage
}
\IEEEauthorblockA{Faculty of Information Technology\\
Monash University, Melbourne, Australia.\\
herath.bandara@monash.edu, christoph.bergmeir@monash.edu, hansika.hewamalage@monash.edu}
}

\maketitle
\copyrightnotice

\begin{abstract}
Generating forecasts for time series with multiple seasonal cycles is an important use-case for many industries nowadays. Accounting for the multi-seasonal patterns becomes necessary to generate more accurate and meaningful forecasts in these contexts. In this paper, we propose Long Short-Term Memory Multi-Seasonal Net (LSTM-MSNet), a decomposition-based, unified prediction framework to forecast time series with multiple seasonal patterns. The current state of the art in this space are typically univariate methods, in which the model parameters of each time series are estimated independently. Consequently, these models are unable to include key patterns and structures that may be shared by a collection of time series. In contrast, LSTM-MSNet is a globally trained Long Short-Term Memory network (LSTM), where a single prediction model is built across all the available time series to exploit the cross-series knowledge in a group of related time series. Furthermore, our methodology combines a series of state-of-the-art multi-seasonal decomposition techniques to supplement the LSTM learning procedure. In our experiments, we are able to show that on datasets from disparate data sources, like e.g. the popular M4 forecasting competition, a decomposition step is beneficial, whereas in the common real-world situation of homogeneous series from a single application, exogenous seasonal variables or no seasonal preprocessing at all are better choices. All options are readily included in the framework and allow us to achieve competitive results for both cases, outperforming many state-of-the-art multi-seasonal forecasting methods.
\end{abstract}


\section{Introduction}
\label{sec:intro}

Time series forecasting has become a key-enabler of modern day business planning by landscaping the short-term, medium-term and long-term goals in an organisation. As such, generating accurate and reliable forecasts is becoming a perpetual endeavour for many organisations, leading to significant savings and cost reductions.
The complex nature of the properties present in a time series, such as seasonality, trend, and level, may bring numerous challenges to produce accurate forecasts. In terms of seasonality, a time series may exhibit complex behaviour such as multiple seasonal patterns, non-integer seasonality, calendar effects, etc.

\begin{figure}[htbp]
\centerline{\includegraphics[width=0.50\textwidth]{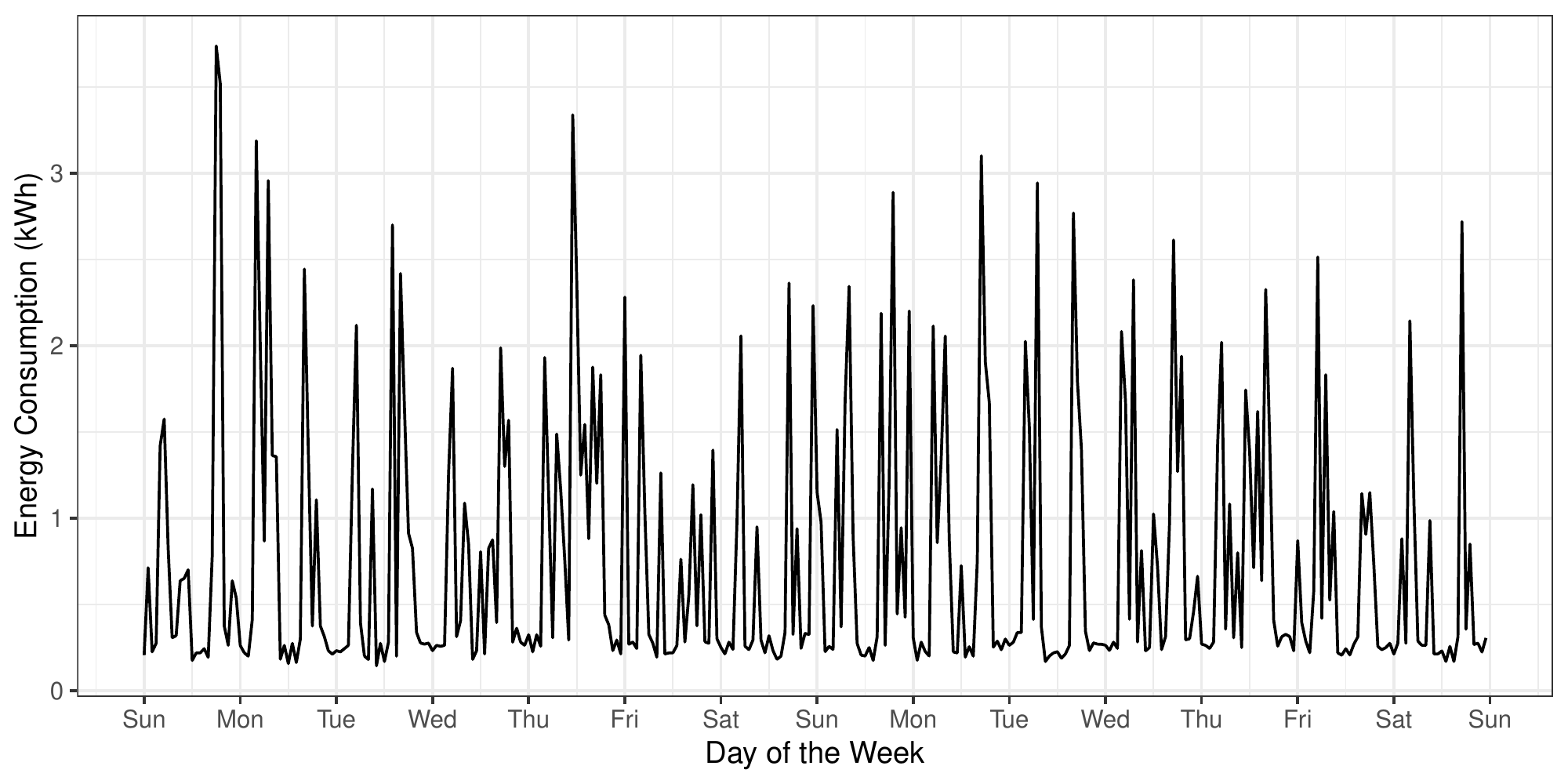}}
\caption{Half-hourly energy consumption of a household over a two weeks period of time, extracted from the AusGrid-Energy Dataset \cite{AusGrid2019-wq}, displaying the inter-day (daily) and intra-day (weekly) seasonal patterns.}
\label{dailyenergyconsumption}
\end{figure}

As sensors and data storage capabilities advance, time series with higher sampling rates (sub-hourly, hourly, daily) are becoming more common in many industries, e.g. in the utility demand industry (electricity and water usage). Fig.~\ref{dailyenergyconsumption} illustrates an example of half-hourly energy consumption of an Australian household that exhibits both daily (period = 48) and weekly (period = 336) seasonal patterns. A longer version of this time series may even exhibit a yearly seasonality (period = 17532), representing seasonal effects such as summer and winter. Particularly in the energy industry, accurate short-term and long-term load forecasting may lead to better demand planning and efficient resource management. In addition to the utility demand industry, the demand variations in the transportation, tourist, and healthcare industries can also be largely influenced by multiple seasonal cycles.

\begin{figure}[htbp]
\centerline{\includegraphics[width=0.49\textwidth]{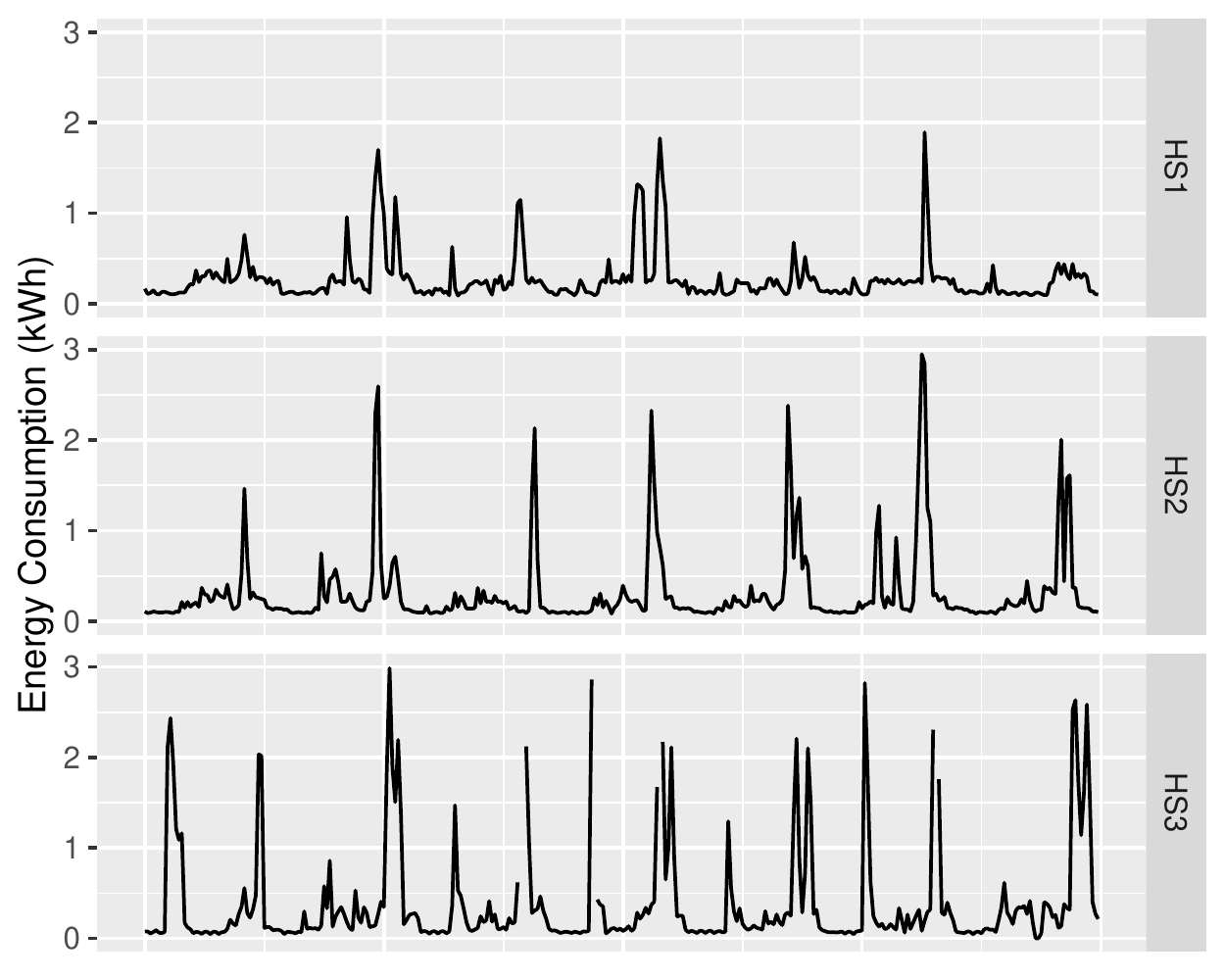}}
\caption{Half-hourly energy consumption fluctuations of three different households in Australia \cite{AusGrid2019-wq} over a time period of one week.}
\label{demandtimeseries}
\end{figure}

The current methods to handle multiple seasonal patterns are mostly statistical forecasting techniques \cite{Lee2011-gy,Box2015-bz} that are univariate. Thus, they treat each time series as an independent sequence of observations, and forecast it in isolation. The univariate time series forecasting is not able to exploit any cross series information available in a set of time series that may be correlated and share a large amount of common features. This is a common characteristic observed in the realm of ``Big Data,'' where often large collections of related time series are available. Examples for these are sales demand of related product assortments in retail, server performance measures in computer centres, household smart meter data, etc. This can be applied to the time series shown in Fig.~\ref{demandtimeseries}, in which these energy consumption patterns of various households can be similar and may share key properties in common. As a result, efforts to build global models across multiple related time series is becoming increasingly popular, and these methods have achieved state-of-the-art performance in recent studies \cite{Borovykh2017-vz, Salinas2017-kx, Wen2017-ky, Bandara2019-iv,Lai2018-zx, Bandara2019-ua}. The recent success is mainly around Recurrent Neural Networks (RNN) and Long Short-Term Memory Networks (LSTM) that are naturally suited in modelling sequence data.

Although several unified models have been proposed to learn better under these circumstances, how to handle multiple-seasonal patterns in a set of time series has not yet been thoroughly studied. Moreover, the competitiveness of such global models highly rely on the  characteristics of the time series. To this end, in this paper, we propose LSTM-MSNet, a novel forecasting framework using LSTMs that effectively accounts for the multiple seasonal periods present in a time series. Following the recent success, our model borrows the strength across a set of related time series to improve the forecast accuracy. This enables our model to untap the common seasonality structures and behaviours available in a collection of time series. As a part of the LSTM-MSNet architecture, we introduce a host of decomposition techniques to supplement the LSTM learning procedure, following the recommendations of \cite{Nelson1999-bd,Zhang2005-pk,Ben_Taieb2012-re,Bandara2019-iv}. Nevertheless, competitiveness of such global models can be affected by the homogeneous characteristics present in the collection of time series~\cite{Bandara2019-iv}. Therefore, LSTM-MSNet introduces two training paradigms to accommodate both homogeneous and inhomogeneous groups of time series. Our model is evaluated using several time series databases, including a competition dataset and real-world datasets, which contain multiple seasonal patterns, exhibiting different levels of seasonal homogeneity. The source code relevant to LSTM-MSNet framework is available at \textit{\url{https://github.com/kasungayan/LSTMMSNet}}

The rest of the paper is organised as follows. In Section~\ref{sec:relatedwork}, we discuss the developments of statistical approaches and neural networks in the multi-seasonal time series forecasting field. Next in Section~\ref{sec:LSTM-MSNet}, we explore the proposed LSTM-MSNet forecasting framework in detail, and outline the key learning paradigms employed in our architecture. Our experimental setup is presented in Section~\ref{sec:eval}, where we demonstrate the results obtained by applying LSTM-MSNet to a variety of time series datasets with multiple-seasonal cycles. Finally, Section~\ref{sec:conclusion} concludes the paper.

\section{Background and related work}
\label{sec:relatedwork}

The traditional approaches to model time series with seasonal cycles are mostly state-of-the-art univariate statistical forecasting methods such as exponential smoothing methods \cite{Hyndman2008-yd} and autoregressive integrated moving-average (ARIMA) models \cite{Box2015-bz}. The basic forms of these algorithms are only suited in modelling a single seasonality, and unable to account for multiple seasonal patterns.

Nonetheless, over the past decade, numerous studies have been conducted to extend the traditional statistical forecasting models to accommodate multiple seasonal patterns \cite{Harvey1997-po, Taylor2003-qp, Gould2008-ug, Taylor2012-nt, De_Livera2011-sg}. An early study developed by Harvey et al. \cite{Harvey1997-po} introduces a model to suit time series with two seasonal periods. Later, Taylor \cite{Taylor2003-qp} adapts the simple Holt-Winters method to capture seasonalities by introducing multiple seasonal components to the linear version of the model. Gould et al. \cite{Gould2008-ug} propose an innovation state space approach to model multiple seasonalities, in which various forms of seasonal patterns can be incorporated, i.e., additive seasonality and multiplicative seasonality. Later, Taylor and Snyder \cite{Taylor2012-nt} overcome the limitations of Gould et al. \cite{Gould2008-ug} by introducing a parsimonious version of the seasonal exponential smoothing approach. However, the majority of these techniques suffer from over-parameterisation, optimisation problems, and are also unable to model complex seasonal patterns in a time series. For more detailed discussions of these weaknesses, we refer to De Livera et al.~\cite{De_Livera2011-sg}. A more flexible and parsimonious version of an innovation state space modelling framework was developed by De Livera et al. \cite{De_Livera2011-sg}, aiming  to address various challenges associated with seasonal time series, such as modelling multiple seasonal periods, non-integer seasonality, and calendar effects. Today, these proposed seasonal models, i.e., BATS and TBATS, are considered state-of-the-art statistical techniques to model time series with multiple seasonal patterns.

Time series decomposition is another popular strategy to handle time series with complex seasonal patterns \cite{Nowicka-Zagrajek2002-go,Lee2011-gy}. Here, the time series is decomposed into a trend, seasonal, and residual component. Each component is modelled separately, so that the model complexity is less than forecasting the original time series as a whole. For example, this approach is applied by Nowicka-Zagrajek and Weron~\cite{Nowicka-Zagrajek2002-go}, where those authors initially decompose time series using a moving average technique. The seasonally adjusted time series is then modelled separately using an ARMA process. Moreover, Lee and Ko~\cite{Lee2011-gy} use a lifting scheme, a different decomposition technique, to separate the original time series at different load frequency levels. Afterwards, individual ARIMA models are built to forecast each decomposed sub series separately.

In parallel to these developments, neural networks (NNs) have been advocated as a strong alternative to traditional statistical forecasting methods in forecasting seasonal time series. The favourable properties towards forecasting, such as universal function approximation \cite{Cybenko1989-fw,Hornik1991-wd}, in theory position NNs as a competitive machine learning approach to model underlying seasonality in a time series. Though early studies postulate the suitability of NNs in modelling seasonal patterns \cite{Zaiyong_Tang1991-wl, Marseguerra1992-fp}, more recent studies advise that deseasonalising the time series prior to modelling is useful to achieve better forecasting accuracy from NNs \cite{Nelson1999-bd,Ben_Taieb2012-re,Zhang2005-pk,Yan2012-dh,Hewamalage2019-il}. Here, deseasonalisation refers to the process of removing the seasonal component from a time series. More specifically, Nelson et al. \cite{Nelson1999-bd} and Ben Taieb et al.~\cite{Ben_Taieb2012-re} empirically show the accuracy gains by including a deseasonalisation process with NNs. Furthermore, Zhang and Qi~\cite{Zhang2005-pk} highlight that NNs are unable to model trend or seasonality directly, thus detrending or deseasonalisation is necessary to produce accurate forecasts with NNs. Meanwhile, Dudek~\cite{Dudek2013-lh} develops a local learning based approach to deal with multiple seasonal cycles. Though this obviates the need of time series decomposition, the local learning procedure that matches similar seasonal patterns in a time series tends to weaken the global generalisability of the model.

More recently, deep neural networks have drawn significant attention among forecasting practitioners. In particular, RNNs and convolutional neural networks (CNN) have exhibited promising results, outperforming many state-of-the-art statistical forecasting methods \cite{Borovykh2017-vz, Salinas2017-kx, Wen2017-ky, Fernando2018-cx, Bandara2019-iv,Lai2018-zx,Han2018-in}. Nevertheless, in spite of the substantial literature available on deep learning in time series forecasting, only few attempts have been undertaken to explicitly handle multiple seasonal patterns in a time series \cite{Lai2018-zx,Bianchi2017-np}. Lai et al. in~\cite{Lai2018-zx} introduce a combination of CNN and RNN architectures to model short and long term dependencies in a time series, and employ a skipped connection architecture to model different seasonal periods. Bianchi et al.~\cite{Bianchi2017-np} implement a seasonal differencing strategy to select the most significant seasonal pattern, i.e., single seasonality present in a time series to forecast the short-term energy load. In order to capture temporal dependencies across both short and long term periods, Fernando et al.~\cite{Fernando2018-cx} develop a  recursive memory network architecture that jointly models both long term and short term relationships. Here, the proposed hierarchical memory structure attempts to jointly model both long term and short term relationships in sequence-to-sequence mapping problems. Moreover, the winning submission of the recently concluded M4 forecasting competition \cite{Makridakis2018-cm}, Exponential Smoothing-Recurrent Neural Network (ES-RNN), uses a hybrid approach to forecast the hourly time series category with two seasonalities. However, the original implementation of ES-RNN restricts the number of seasonalities to two, and also due to the limitations of the underlying models (Holt-Winters) that operate in this approach, the ES-RNN is not suitable to handle long term seasonalities in a time series (e.g., yearly seasonality in an hourly time series) \cite{Smyl2019-cb}.

\section{LSTM-MSNet Framework}
\label{sec:LSTM-MSNet}
In this section, we first formally define the problem of forecasting with multiple seasonal patterns, and then discuss the components of the proposed LSTM-MSNet architecture.

\subsection{Problem Statement}
\label{sec:problem}
Let $i\ \in \{1, 2, ..., n\}$ be the $i$th time series from $n$ time series in our database. The past observations of the time series $i$ are given by $X_i = \{x_1, x_2, ..., x_K\} \in {\Bbb R^{K_{i}}}$, where $K_i$ represents the length of the time series $i$. We introduce the seasonality periods of time series $i$ as $S_i = \{s_1, s_2, ..., s_P\} \in {\Bbb R^{P}}$, where $P$ is the highest seasonal period present in the time series $i$. The primary objective of this study is to develop a global prediction model $f$, which uses previous observations of all the time series, i.e., $X = \{X_1, X_2, ..., X_n\}$ to forecast $M$ number of future data points $i$, i.e., $X^M_{i} = \{x_t, x_{t+1}, ..., x_{t+M}$\}, while accounting for all the available seasonal periods $S = \{S_1, S_2, ..., S_P\} \in {\Bbb R^{n \times P}}$ present in the time series. Here, $M$ is the intended forecasting horizon of time series $i$. The model $f$ can be defined as follows:

\begin{equation}
  X^M_{i} = f(X,S,\theta)
  \label{eq0}
\end{equation}
Here, $\theta$ are the model parameters of our LSTM-MSNet prediction model.

LSTM-MSNet is a forecasting framework designed to forecast time series with multiple seasonal patterns. The architecture of LSTM-MSNet is a fusion of statistical decomposition techniques and recurrent neural networks. The LSTM-MSNet has three layers, namely: 1) the pre-processing layer, which consists of a normalisation and variance stabilising phase, and a seasonal decomposition phase, 2) the recurrent layer, which consists of an LSTM based stacking architecture to train the network, and 3) a post-processing layer to denormalise and reseasonalise the time series to derive the final forecasts. The proposed framework can be used with any RNN variant such as LSTMs, Gated Recurrent Units (GRUs), and others. In this paper, we select LSTMs, a promising RNN variant, as our primary network training module. In the following sections, we discuss each layer of the LSTM-MSNet in detail.

\subsection{Normalisation and Variance Stabilisation Layer}
The proposed LSTM-MSNet is a global model that is built across a group of time series. Therefore, performing a data normalisation strategy becomes necessary as in a collection of time series, each time series may contain observations with different value ranges. Hence, we use the mean-scale transformation strategy, which uses the mean of a time series as the scaling factor. This scaling strategy can be defined as follows:

\begin{equation}
x_{i, normalised} = \frac{x_i}{\frac{1}{k}\sum_{t=1}^{k}{x_{i,t}}}
\label{meanscale}
\end{equation}

Here, $x_{i, normalised}$ represents the normalised observation, and $k$ represents the number of observations of time series $i$.

After normalising the time series, we stabilise the variance in the group of time series by transforming each time series to a logarithmic scale. Apart from the variance stabilisation, the log transformation also enables the conversion of the seasonality form in a given time series to an additive form. This is a necessary requirement for additive time series decomposition techniques employed in our decomposition layer. The transformation can be defined in the following way:

\begin{equation}
X_{i, logscaled} = \begin{cases}
  \log(X_i), & min(X)>0;\\
  \log(X_i+1), & min(X)=0;\\
\end{cases}
\label{logscale}
\end{equation}

Here, $X$ denotes a time series, and $X_{i, logscaled}$ is the corresponding log transformed time series $i$.

\subsection{Seasonal Decomposition}

As highlighted in Section~\ref{sec:relatedwork}, when modelling seasonal time series with NNs, many studies suggest applying a prior seasonal adjustment, i.e., deseasonalisation to the time series \cite{Nelson1999-bd,Ben_Taieb2012-re,Zhang2005-pk}. The main intention of this approach is to minimise the complexity of the original time series, and thereby reducing the subsequent effort of the NN’s learning process. In line with these recommendations, LSTM-MSNet initially uses a deseasonalisation strategy to detach the multi-seasonal components from a time series. Here, seasonal components refer to the repeating patterns that exist in a time series and that may change slowly over time \cite{Dokumentov2018-us}. To accommodate this, we use a series of statistical decomposition techniques that support separating multi-seasonal patterns in a time series.  We also configure these methods to extract various forms of seasonality, i.e., deterministic and stochastic seasonality to assess their sensitivity towards the forecast accuracy. Next, we briefly describe the different types of decomposition techniques used in our study. An overview of the methods is given in Table~\ref{tab:decompmethods}.

\subsubsection{Multiple STL Decomposition (MSTL)}

MSTL extends the original version of Seasonal-Trend Decomposition (STL) \cite{Cleveland1990-rc}, to allow for decomposition of a time series with multiple seasonal cycles. The STL method additively decomposes a time series into trend, seasonal, and remainder components. In other words, the original series can be reconstructed by summing the decomposed parts of the time series. The additive decomposition can be formulated as follows:

\begin{equation}
x_t = \hat{S}_t + \hat{T}_t  + \hat{R}_t
\label{additive}
\end{equation}

Here, $x_t$ represents the observation at time $t$, and $\hat{S}_t $, $\hat{T}_t$, $\hat{R}_t$ refers to the seasonal, trend, and the remainder components of the observation, respectively.

In MSTL, the STL procedure is used iteratively to estimate the multiple seasonal components in a time series. So, the original version of Equation~\ref{additive} can be extended to reflect the decomposition of MSTL as follows:

\begin{equation}
x_t = \hat{S^1}_t + \hat{S^2}_t + ... + \hat{S^n}_t + \hat{T}_t  + \hat{R}_t
\label{multiadditive}
\end{equation}

Here, $n$ denotes the number of distinct seasonal patterns decomposed by the MSTL. In our study, we use the R \cite{Rlanguage2013} implementation of the MSTL algorithm, \verb|mstl|, from the \verb|forecast| package \cite{Hyndman2015-vm,Khandakar2008-hd}. MSTL also supports controlling the smoothness of the change of seasonal components extracted from the time series, i.e., configuring the \verb|s.window| parameter. For example, by adjusting the \verb|s.window| parameter to ``periodic'', the MSTL decomposition limits the change in the seasonal components to zero. This enables us to separate the deterministic seasonality from a time series. In Table~\ref{tab:decompmethods}, we give the two \verb|s.window| parameter values used in our experiments.

\subsubsection{Seasonal-Trend decomposition by Regression (STR)}

STR is a regression based decomposition technique introduced by Dokumentov et al.~\cite{Dokumentov2015-fy}. The division is additive, hence the decomposition accords with Equation~\ref{multiadditive}. In contrast to STL, STR is capable of incorporating multiple external regressors to the decomposition procedure, while allowing to account for external factors that may influence the seasonal patterns in a time series. However, to make our comparisons unbiased, we use STR in the default mode, without including any exogenous regressors. In R, the STR algorithm is available through the \verb|AutoSTR| function from the \verb|stR| package \cite{Dokumentov2018-us}.

\subsubsection{Trigonometric, Box-Cox, ARMA, Trend, Seasonal (TBATS)}

As highlighted in Section~\ref{sec:relatedwork}, the TBATS model was developed to handle complex seasonal patterns present in a time series \cite{De_Livera2011-sg}. This method is currently established as a state-of-the-art technique to forecast time series with multiple seasonal cycles. Particularly, the inclusion of trigonometric expression terms has enabled TBATS to identify sophisticated seasonal terms in a time series (for details see Livera et al. \cite{De_Livera2011-sg})).

In our seasonal decomposition step, we use TBATS as a deseasonalisation technique to extract the relevant seasonal components of a time series. We perform the seasonal extraction after fitting the TBATS model using the \verb|tbats| function provided by the \verb|forecast| package \cite{Hyndman2015-vm,Khandakar2008-hd} in R.

\subsubsection{Prophet}
Prophet is an automated forecasting framework developed by Taylor and Letham~\cite{Taylor2017-lw}. The main aim of this framework is to address the challenges involved in forecasting at Facebook, the employer of those authors at that time. The challenges include the task of forecasting time series with multiple seasonal cycles. The underlying model of Prophet uses an additive decomposition layer similar to Equation~\ref{multiadditive}. However, this division introduces an additional term to model holidays as seasonal covariates. After including the holiday terms, Equation~\ref{multiadditive} can be rewritten as follows:

\begin{equation}
x_t = \hat{S^1}_t + \hat{S^2}_t + ... + \hat{S^n}_t + \hat{T}_t  + \hat{R}_t + \hat{H}_t
\label{multiadditive2}
\end{equation}
 
Here, $\hat{H}_t$ denotes the holiday covariates in the model that represent the effects of holidays. Likewise in TBATS, we use Prophet in the Decomposition layer to obtain the multiple seasonal components present in a time series. We achieve this by applying the Prophet algorithm available through the \verb|prophet| package in R \cite{Taylor2018-wb}.

\subsubsection{Fourier Transformation}

Fourier terms are a flexible approach to model periodic effects in a time series \cite{Harvey1993-sg}. For example, let $x_t$ be an observation of time series $X$ at time $t$. The seasonal terms relevant to $x_t$ can be approximated by Fourier terms as follows:

\begin{equation}
\small
\sin\left(\frac{2\pi kt}{s_1}\right),
  \cos\left(\frac{2\pi kt}{s_1}\right), ... , \sin\left(\frac{2\pi kt}{s_n}\right),
  \cos\left(\frac{2\pi kt}{s_n}\right)
\label{fourierterms}
\end{equation}

Here, $s_n$ refers to the $n$th seasonal periodicity in the time series. Thereby, we can define an amount of $n$ seasonal periodicities available in a time series. The parameter $k$ in Equation~\ref{fourierterms} is the number of $sin$, $cos$ pairs used for the transformation process. This essentially controls the momentum of the seasonality, where a higher $k$ allows to represent a seasonal pattern that changes more quickly, compared to a lower $k$. In our case, for each seasonal periodicity in the time series, a separate $k$ must be introduced. We generate these Fourier terms using the \verb|fourier| function available in the \verb|forecast| package. In our experiments, we use a parameter grid, which ranges from $k = 1$ to $k = s/2$, to determine the optimal $k$ values in Fourier terms. Moreover, we consider $k = 1$ (with least number of $k$) as a special use case in Fourier terms and report separately as a variant of LSTM-MSNet.

The overall summary of the aforementioned methods is shown in Table~\ref{tab:decompmethods}. Here, the \textit{Package} column provides a reference to the software implementation used in our experiments. The table furthermore indicates the type of seasonalities extracted by each method.

\begin{table}
\caption{Summary of techniques used for multi-seasonal decomposition}
\centering
\resizebox{\columnwidth}{!}{%
\begin{tabular}{lccc}
\hline
Technique &Package &Deterministic &Stochastic \\
\hline
MSTL \scriptsize{(s.window = ``periodic'')} &forecast \scriptsize{\cite{Hyndman2015-vm}} &\color{green}\checkmark &\color{red}\xmark\\
MSTL \scriptsize{(s.window = ``7'')} &forecast \scriptsize{\cite{Hyndman2015-vm}} &\color{green}\checkmark &\color{green}\checkmark\\
AutoSTR &stR \scriptsize{\cite{Dokumentov2018-us}} &\color{green}\checkmark &\color{green}\checkmark\\
TBATS &forecast \scriptsize{\cite{Hyndman2015-vm}} &\color{green}\checkmark &\color{green}\checkmark\\
Prophet &prophet \scriptsize{\cite{Taylor2018-wb}} &\color{green}\checkmark &\color{green}\checkmark\\                      
\hline
\end{tabular}
}
\label{tab:decompmethods}
\end{table}

\subsection{Recurrent Layer}
The second layer, the Recurrent Layer, is the primary prediction module of LSTM-MSNet, equipped with LSTMs. RNNs, and in particular LSTMs, have been embraced by many fields that involve sequence modelling tasks, such as Natural Language Processing \cite{Mikolov2010-rb}, speech recognition \cite{Graves2013-cu}, image generation \cite{Gregor2015-rt}, and more recently have received a great amount of attention in time series research \cite{Zimmermann2012-cp,Salinas2017-kx, Wen2017-ky, Bandara2019-iv,Hewamalage2019-il}. 

In the LSTM, the gating mechanism together with the self-contained memory cell enables the network to capture non-linear long-term temporal dependencies in a sequence. We configure the input and forget gates of the LSTM network to include the previous state of the memory cell ($C_{t-1}$). This configuration is also known as ``LSTM with peephole connections'', in which the hidden state ($h_t$) at time $t$ can be computed from the following equations:

\begin{equation}
\label{input}
 i_t              = {\sigma}(W_{i}{\cdot}h_{t-1} + U_{i}{\cdot}x_{t} + P_{i}{\cdot}C_{t-1} + b_{i})
\end{equation}
\begin{equation}
\label{forget}
 f_t              = {\sigma}(W_{f}{\cdot}h_{t-1} + U_{f}{\cdot}x_{t} + P_{f}{\cdot}C_{t-1}+ b_{f})
\end{equation}
\begin{equation}
\label{candidate}
 \tilde{C_{t}}    = {\tanh}(W_{c}{\cdot}h_{t-1} + U_{c}{\cdot}x_{t} + b_{c})
\end{equation}
\begin{equation}
\label{cell}
  C_t              = f_{t}\, {\odot} \,C_{t-1} \ + \, i_{t}\, {\odot}\, \tilde{C_{t}}
\end{equation}
\begin{equation}
\label{output}
  o_t              = {\sigma}(W_{o}{\cdot}h_{t-1} + U_{o}{\cdot}x_{t} + P_{o}{\cdot}C_{t}+ b_{o})
\end{equation}
\begin{equation}
\label{state}
  h_t 			    = o_t \,{\odot}\,{\phi}(C_t)
\end{equation}

Here, $W_i$, $W_f$, $W_o$, and  $W_c$ represent the weight matrices of input gate, forget gate, output gate, and memory cell gates respectively, while $x_t$ is the input at time $t$. Also, $U_i$, $U_f$, $U_o$, and $U_c$ denote the corresponding input weight matrices, and $P_{i}$, $P_{f}$, $P_{o}$ are the respective peephole weight matrices. The biases of the gates are represented by $b_i$, $b_f$, $b_o$, and  $b_c$. $\tilde{C_{t}}$  refers to the candidate cell state, which is used to update the state of the original memory cell $C_{t}$ (see Equation~\ref{cell}). In these equations, $\odot$ represents the element-wise multiplication operation, $\sigma$ represents the logistic sigmoid activation function, and $\phi$ stands for the hyperbolic tangent function, i.e., \textit{tanh}. Furthermore, we use \textit{tanh} as our hidden update activation function in Equation~\ref{candidate}.

\subsubsection{Moving Window Transformation}
As a preprocessing step, we transform the past observations of time series ($X_i$) into multiple pairs of input and output frames using a Moving Window (MW) strategy. Later, these frames are used as the primary training source of LSTM-MSNet.

\begin{figure}[htbp]
\centerline{\includegraphics[width=0.50\textwidth]{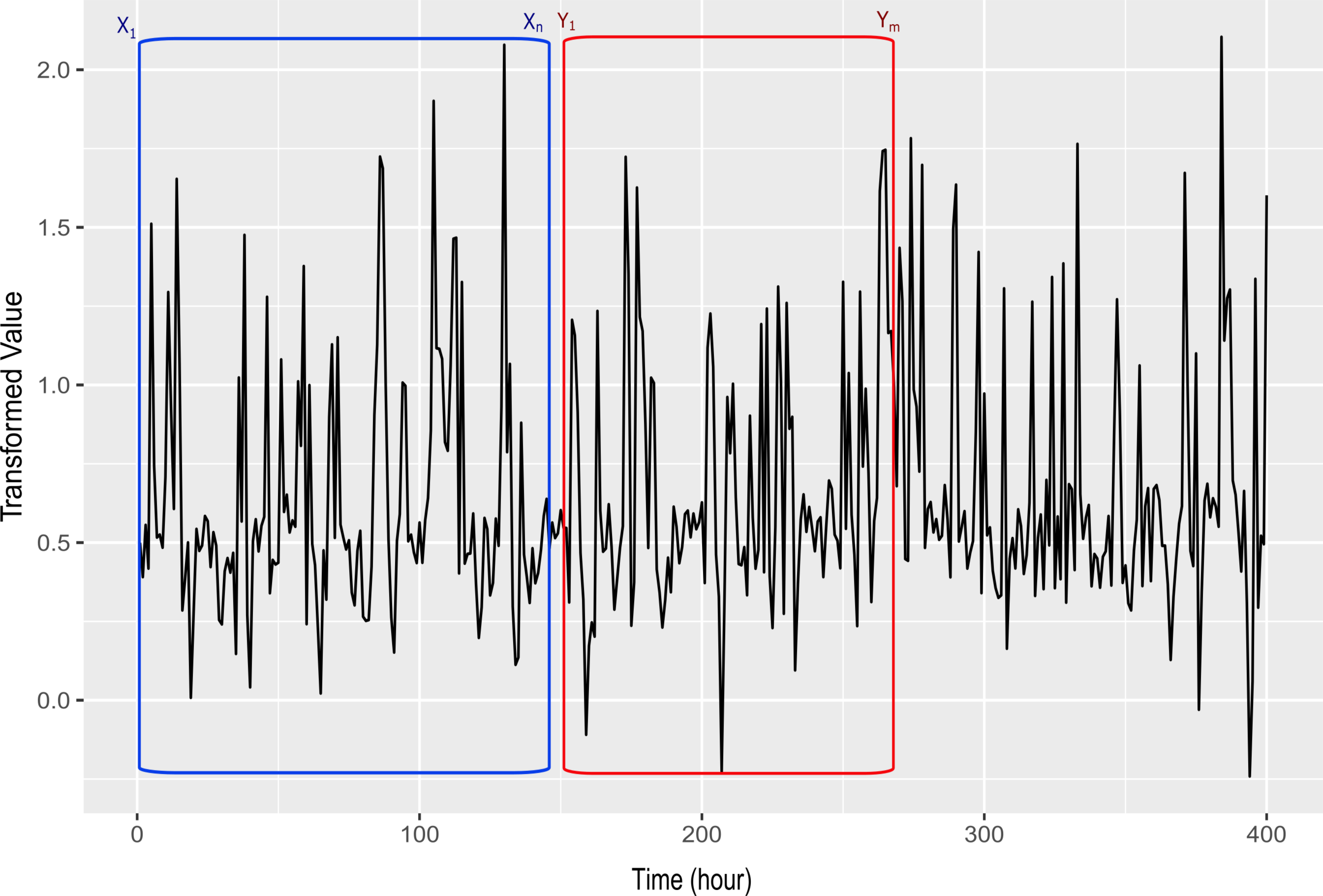}}
\caption{An example of applying the moving window approach to a preprocessed and seasonally adjusted time series $X_i$, i.e., after the normalisation \& variance stabilisation and decomposition layers. Here, $\{x_1, x_2, ..., x_n\}$ represents the input window, and $\{y_1, y_2, ..., y_m\}$ corresponds to the output window.}
\label{mw}
\end{figure}

In summary, the MW strategy converts a time series $X_i$ of length $K$ into $(K -n -m)$ records, where each record has an amount of $(m + n)$ observations. Here, $m$ refers to the length of the output window, and $n$ is the length of the input window. These frames are generated according to the Multi-Input Multi-Output (MIMO) principle used in multi-step forecasting, which directly predicts all the future observations up to the intended forecasting horizon $X^M_{i}$. Training the NNs in this way has the advantage of avoiding the potential error accumulation at each forecasting step \cite{Ben_Taieb2012-re, Wen2017-ky,Ben_Taieb2016-il}. Therefore, we choose the size of the output window $m$ equivalent to the length of the intended forecast horizon $M$ ($m=M$, the list of $M$ values used for the benchmark datasets are summarised in Table II, under column $M$). Fig.~\ref{mw} illustrates an example of applying the MW approach to the hourly time series T48 of the M4 dataset.

To train the LSTM-MSNet, we use $(K -m)$ many observations from time series $X_i$ and reserve the last output window for network validation and model hyper-parameter tuning.

\subsubsection{Training Paradigms}
\label{sec:tp}

\begin{figure*}[htp]
\subfloat[The proposed DS training paradigm used to train the LSTM-MSNet]{%
  \centerline{\includegraphics[width=0.80\textwidth]{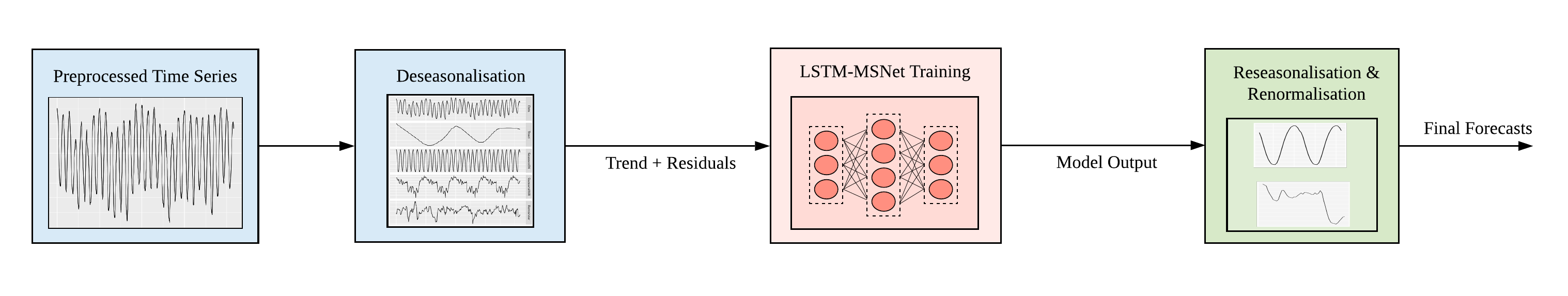}}
}
\vspace{4mm}
\subfloat[The proposed SE training paradigm used to train the LSTM-MSNet]{%
  \centerline{\includegraphics[width=0.80\textwidth]{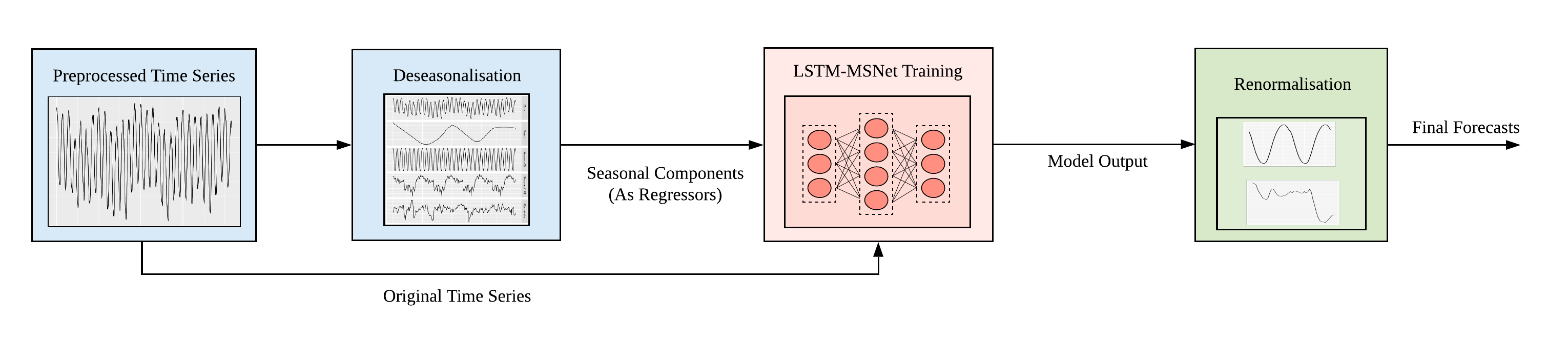}}
}
\caption{An overview of the proposed LSTM-MSNet training paradigms. In the DS approach, deseasonalised time series are used to train the LSTM-MSNet. Here, a reseasonalisation phase is required as the target MW patches are seasonally adjusted. Whereas in the SE approach, the seasonal values extracted from the deseasonalisation phase are employed as exogenous variables, along with the original time series to train the LSTM-MSNet. Here a reseasonalisation phase is not required as the target MW patches contain the original distribution of the time series.}
\label{lstm-paradigms}
\end{figure*}

In this study, we propose to use the output of the decomposition layer in two different ways. These paradigms can be distinguished by the time series components used in the MW process, and later in the LSTM-MSNet training procedure. In the following, we provide a short overview of these two paradigms shown in Fig.~\ref{lstm-paradigms} .

\paragraph{Deseasonalised Approach (DS)}
 
This approach uses seasonally adjusted time series as MW patches to train the LSTM-MSNet. Since the seasonal components are not included in DS for the training procedure, a reseasonalisation technique is later introduced in the Post-processing layer of LSTM-MSNet to ascertain the corresponding multiple seasonal components of the time series. 

\paragraph{Seasonal Exogenous Approach (SE)}

This second approach uses the output of the pre-processing layer, together with the seasonal components extracted from the multi-seasonal decomposition as external variables. Here, in addition to the normalised time series (without the deseasonalisation phase), the seasonal components relevant to the last observation of the input window are used as exogenous variables in each input window. As the original components of the time series are used in the training phase of SE, the LSTM-MSNet is expected to forecast all the components of a time series, including the relevant multi-seasonal patterns. Therefore, a reseasonalisation stage is not required by SE.

In summary, DS supplements the LSTM-MSNet by excluding the seasonal factors in the LSTM-MSNet training procedure. This essentially minimises the overall training complexity of the LSTM-MSNet. In contrast, SE supplements LSTM-MSNet in the form of exogenous variables that assist modelling the seasonal trajectories of a time series.

The DS paradigm can be seen as a boosting ensemble technique \cite{Schapire2003-xb}, where the deseasonalisation process is a weak base learner that is subsequently supplemented by the LSTM, which is trained on the remainder of the base learner. Here, the complexity of the base learner, i.e., the different deseasonalisation techniques, can affect the subsequent LSTM training procedure and may lead to different results. Thereby, the suitability of the base learner as a good pre-processing technique is more important than its overall forecasting accuracy. This is a conclusion that can also be drawn from the poor performance of our submission to the M4 competition \cite{kasunbM4-sf}, where we used ETS \cite{Hyndman2008-yd} and TBATS \cite{De_Livera2011-sg}, which can be considered strong base learners, to fit the time series. Subsequently, an LSTM was trained on the residuals of those models. Therefore, in the DS approach, the equilibrium between the base learner and the LSTM determines the final performance of the models, so that it seems worthwhile to test different such base learners.

On the other hand, the SE paradigm can be seen as an encompassing version of the DS case, where as a special case the LSTM could ``learn'' to subtract the exogenous input from the time series. However, during the LSTM learning phase, the LSTM only sees a particular input window at a time. This input, depending on how long a seasonal period is and where in the period of seasonality the current input comes from, may look dramatically different. So, learning such seemingly simple relationships from a limited set of input data may be in fact a difficult task for the LSTM. Consequently, our work empirically examines the capacity of the LSTM to learn in different practical situations.

\subsubsection{LSTM Learning Scheme}

\begin{figure}[htbp]
\centerline{\includegraphics[width=0.48\textwidth]{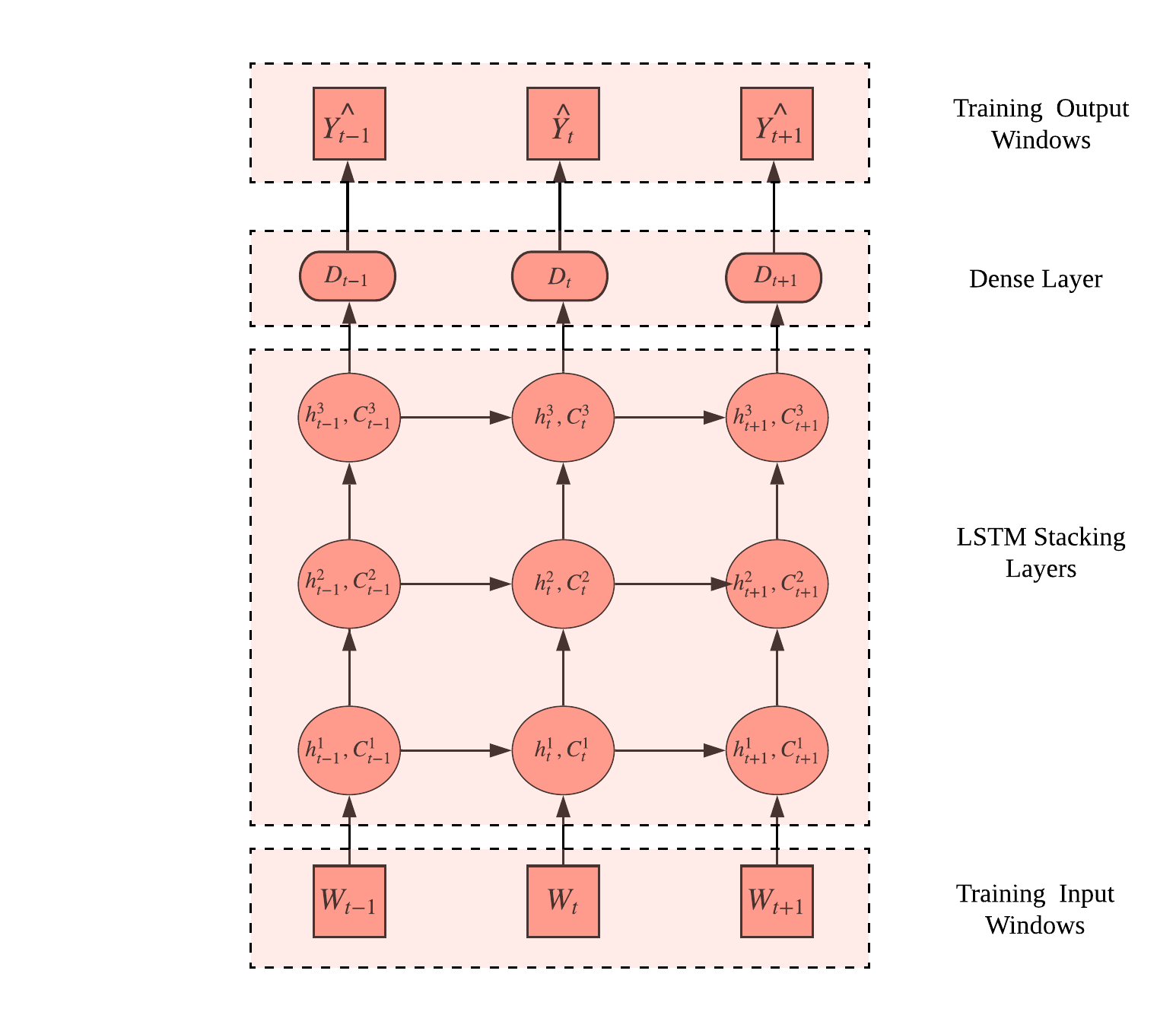}}
\caption{The unrolled representation of a peephole connected LSTM in time, with the hidden state ($h_t$) and memory cell ($C_t$). In this architecture, $h_t$ expects to capture the short-term dependencies in a sequence, while $C_t$ accounts for the long-term dependencies. Here, the input window, dense layer, and projected LSTM output at time step $t$ are denoted by $W_t$, $D_t$, and $\hat{Y_{t}}$ respectively.}
\label{lstm-schemes}
\end{figure}

As highlighted earlier, we use the past observations of time series $X_i$, in the form of input and output windows to train the LSTM-MSNet. In our work, we follow the LSTM design guidelines recommended by Hewamalage et al.~\cite{Hewamalage2019-il}. Fig.~\ref{lstm-schemes} illustrates the primary LSTM learning architecture of LSTM-MSNet. This consists of four components, namely: Training input window layer, LSTM stacking layer, Dense layer and Training output window layer. Here, $W_{t}\in{\Bbb R^{n}}$ represents the input window at time step $t$. Also, the projected cell output of the LSTM at time step $t$ is represented by $\hat{Y_{t}}\in{\Bbb R^{m}}$. Here $m$ represents the size of the output window, which is identical to the forecasting horizon $M$. Moreover, the hidden and the cell states of the LSTM are denoted by $h_t$ and $C_t$. Here, $h_t$ along with the $C_t$ cell state provides a notion of memory to our learning scheme, while accounting for the dependencies that are longer than a given training input window. We use an affine neural layer (a fully connected layer; $D_t$), excluding the bias component to map each LSTM cell output $h_t$ to the dimension of the output window $m$.

Before feeding these windows to the network for training, each input and output window is subjected to a local normalisation process to avoid possible network saturation effects caused by the bounds of the network activation functions \cite{Bandara2019-iv}. In the DS approach, we use the trend component of the last value of the input window as a local normalisation factor, whereas in SE we use the mean value of each input window as the normalisation factor. Afterwards, these factors are subtracted from each data point in the corresponding input and output window. The above LSTM learning scheme is implemented using TensorFlow \cite{Abadi2016-rr}.

\subsubsection{Loss Function}

We use the L1-norm, as the primary learning objective function, which essentially minimises the absolute differences between the target values and the estimated values. This has the advantage of being more robust to anomalies in the time series. The L1-loss is given by:

\begin{equation}
\mathcal{L}_{1} = \sum_{t \varepsilon \Omega_{Train}}^{}\left | {Y_{t}} - \hat{Y_{t}} \right | + \underbrace{\psi\sum_{i=1}^{p} w_{i}^2}_\text{L2 regularisation}
\label{loss-function}
\end{equation}

Here, ${Y_{t}}\in{\Bbb R^{m}}$ refers to the actual observations of values in the output window at time step $t$. The cell output of the LSTM at time step $t$ is defined by $\hat{Y_{t}}$. Also, $\Omega_{Train}$ is the set of time steps used for training. We include an L2-regularisation term to minimise possible overfitting of the network. In Equation~\ref{loss-function}, $\psi$ is the regularisation parameter, $w_{i}$ refers to the network weights and $p$ is the number of weights in the network.

\subsection{Post-processing Layer}

The reseasonalisation and renormalisation is the main component of the post processing layer in LSTM-MSNet. Here, in the reseasonalisation stage, the relevant seasonal components of the time series are added to the forecasts generated by the LSTM. This is computed by repeating the last seasonal components of the time series to the intended forecast horizon. As outlined in Section~\ref{sec:tp}, SE does not require this phase. Next, in the renormalisation phase, the generated forecasts are back-transformed to their original scale by adding back the corresponding local normalisation factor, and taking the exponent of the values. The final forecasts are obtained by multiplying this vector by the scaling factor used for the normalisation process.

\section{Experiments}
\label{sec:eval}
In this section, we evaluate the proposed variants of the LSTM-MSNet framework on three time series datasets. First, we describe the datasets, error metrics, hyper-parameter selection method, and benchmarks used in our experimental setup. Then, we provide a detailed analysis of the results obtained.

\subsection{Datasets}
\label{sec:datasets}
We use three benchmark time series datasets which present multiple seasonal cycles. Following is a brief overview of these datasets:
\begin{itemize}
\item M4-Hourly Dataset~\cite{Makridakis2018-cm}: Hourly dataset from the M4 forecasting competition.
\item AusGrid-Energy Dataset~\cite{AusGrid2019-wq}: Half-hourly dataset, representing energy consumption of 300 households in Australia. We select general consumption (GC letter code in the dataset) as the primary measure of energy consumption in households.  Firstly, we extract a subset of 3 months of half-hourly data (2012 July - 2012 October). Then, to evaluate LSTM-MSNet on multiple seasonal patterns, we aggregate the original half-hourly time series to hourly time series and extend the extraction to 2 years (2010 July - 2012 July), considering three seasonalities: daily, weekly, and yearly.
\item Traffic Dataset~\cite{traffic2019-sf}: A collection of hourly time series, representing the traffic occupancy rate of different car lanes of San Francisco bay area freeways.
\end{itemize}

Table \ref{tab:datasummary} summarises statistics of the datasets used in our experiments. Here, $N$ denotes the number of time series, $K$ denotes the length of each time series, $T$ denotes the sampling rate of the time series, $S$ represents the different seasonal cycles present in the time series, and $M$ is the relevant forecast horizon. Moreover, we choose the size of the input window $n$ equivalent to $M*1.25$, following the heuristic proposed in \cite{Bandara2019-iv,Hewamalage2019-il}.

\begin{table}
\caption{dataset statistics}
\centering
\resizebox{\columnwidth}{!}{%
\begin{tabular}{lccccc}
	\toprule
	Data set           					&N 			&K    	&T   			&S    				&M  \\ \hline
	M4-Hourly Dataset  					&414		&700	&hourly		   &(24, 168)			&48	\\
	AusGrid-Energy Dataset (3 Months)	&300		&4704	&half-hourly   &(48, 336)			&96	 \\
	AusGrid-Energy Dataset (2 Years)	&300		&17600	&hourly  	   &(24, 168, 8766)		&24	 \\
	Traffic								&963		&700	&hourly		   &(24, 168)			&24 \\
	 \hline
\end{tabular}
}
\label{tab:datasummary}
\end{table}

\begin{figure*}[htbp]
\centerline{\includegraphics[width=0.75\textwidth]{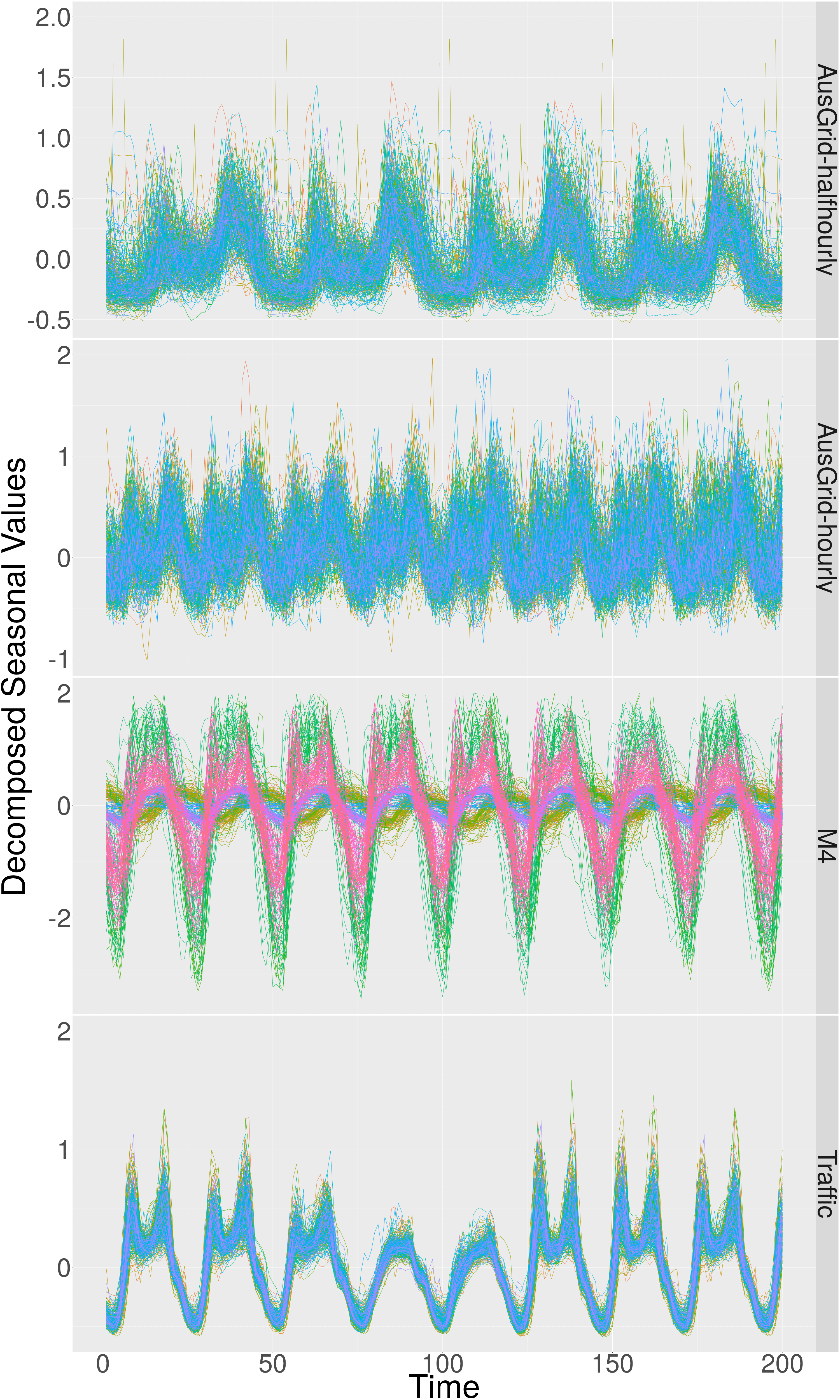}}
\caption{The seasonal components' distributions of the sum of multiple seasonalities extracted from the AusGrid-Energy (half hourly), AusGrid-Energy (hourly), M4 and the Traffic datasets, by applying the MSTL decomposition technique to the initial 200 data points of each time series.}
\label{seasonalpatterns}
\end{figure*}

We plot the seasonal components of our benchmark datasets to investigate the diversity of their seasonal distributions. For simplicity, we use \textit{MSTL} as the primary decomposition technique to extract the multiple seasonal components from a time series. From Fig.~\ref{seasonalpatterns}, it is evident that there exists a high variation of seasonality among the time series in the M4 dataset. This can be attributed to a less homogeneous nature of the time series and different start/end calendar dates. On the other hand, it is clear that the distribution of the seasonal components is similar among the time series in the AusGrid-Energy and Traffic datasets, which shows less variation compared to the M4 dataset, as the time series are homogeneous in the sense that they are all related to household energy consumption, and follow identical calendar dates. This seasonal diversity present in our benchmark datasets enables us to assess the robustness of the LSTM-MSNet framework under different seasonality conditions, i.e., inhomogeneous and homogeneous seasonalities.

\subsection{Error Metrics}
To assess the accuracy of LSTM-MSNet against the benchmarks, we use two evaluation metrics commonly found in the forecasting literature, namely the symmetric Mean Absolute Percentage Error (sMAPE) and the Mean Absolute Scaled Error (MASE) \cite{Hyndman2006-ue}. The sMAPE and MASE are defined as follows:

\begin{equation}
\text{sMAPE} = \frac{2}{m}\sum_{t=1}^{m}\left(\frac{\left|F_t - Y_t\right|}{\left| F_t\right| + \left| Y_t\right|} \right)
\label{smape}
\end{equation}

\begin{equation}
\text{MASE} = \frac{1}{m}\frac{\sum_{t=1}^{m}|F_t - Y_t|}{\frac{1}{n-S}\sum_{t=S+1}^{n}|Y_{t}-Y_{t-S}|}
\label{mase}
\end{equation}

Here, $Y_t$ represents the observation at time $t$, and $F_t$ is the generated forecast. Also, $m$ denotes the number of data points in the test set and $n$ is the number of observations in the training set of a time series. We define $S$, as the frequency of the highest available seasonality in the given time series (e.g., S = 168, S = 336, S = 8766, and S = 168 for the M4-Hourly, AusGrid-Energy, and Traffic datasets respectively). Furthermore, when calculating the sMAPE values for the energy datasets and the traffic dataset, to avoid problems for zero forecasts and actual observations, we add a constant term of $\epsilon = 1$ to the denominator of Equation~\ref{smape}. To provide a broader overview of the error distributions, we compute the mean and median of these primary error measures. This includes, mean of the sMAPEs (Mean sMAPE), median of the sMAPEs (Median sMAPE), mean of the MASEs (Mean MASE), and median MASEs (Median MASE).

\subsection{Statistical tests of the results}

We use the non-parametric Friedman rank-sum test to assess the statistically significant differences within the compared forecasting methods. Also, Hochberg's post-hoc procedure is used to further examine these differences~\cite{Garcia2010-jx}\footnote{More information can be found on the thematic web site of SCI2S about \emph{Statistical Inference in Computational Intelligence and Data Mining \url{http://sci2s.ugr.es/sicidm}}}. The sMAPE error measure is used to perform the statistical testing, with a significance level of $\alpha$ = 0.05.

\subsection{Hyper-parameter Tuning and Optimisation}

The recurrent layer in LSTM-MSNet has various hyper-parameters. This includes LSTM cell dimension, number of epochs, hidden-layers, mini-batch-size and model regularisation terms. In the optimisation framework, we use COntinuous COin Betting \cite{Orabona2017-cd} as our primary learning algorithm to train the network, which does not require the tuning of learning rate. In this way, we minimise the overall amount of hyper-parameters to be tuned in the network.

Furthermore, we automate the hyper-parameter selection process by employing a Bayesian global optimisation methodology  that autonomously discovers the optimal set of parameters of an unknown function. Compared to other parameter optimisation selection techniques, such as Random Search and Grid Search, the Bayesian optimisation strategy is considered as a more systematic and/or more efficient approach. This is because, when determining the current pool of potential hyper-parameters for validation, it takes into account the previously visited hyper-parameter values in a non-trivial way~\cite{Snoek2012-ko}. In our experiments, we use the python implementation of the sequential model-based algorithm configuration (SMAC) \cite{Hutter_undated-rw} that implements a version of the Bayesian hyper-parameter optimisation process.
Table \ref{tab:parametergrid} summarises the bounds of the hyper-parameter values used throughout the LSTM-MSNet learning process, represented by the respective minimum and maximum values.

\begin{table}
\caption{Recurrent layer hyper-parameter grid}
\begin{center}
\begin{tabular}{lcc}
	\toprule
	Model Parameter            			&Minimum value 			&Maximum value\\ \hline
	LSTM-cell-dimension  				&20						&50	\\
	Mini-batch-size						&20						&80	\\
	Epoch-size							&2						&10 \\
	Maximum-epochs						&10						&40	\\
	Hidden Layers						&1						&2\\
	Gaussian-noise-injection			&$10^{-4}$				&$8 \cdot 10^{-4}$
\\
	L2-regularisation-weight			&$10^{-4}$				&$8 \cdot 10^{-4}$\\ \hline
\end{tabular}
\label{tab:parametergrid}
\end{center}
\end{table}

\subsection{Benchmarks and LSTM-MSNet variants}
We compare our developments against a collection of current state-of-the-art techniques in forecasting with multiple seasonal cycles. This includes \textit{Tbats} \cite{De_Livera2011-sg}, \textit{Prophet} \cite{Taylor2017-lw}, and \textit{FFORMA} \cite{Montero-Manso2018-pd}. We also use two variants of \textit{Dynamic-Harmonic-Regression} \cite{Hyndman2015-vm} as the benchmarks. In our experiments, Dynamic-Harmonic-Regression (T) represents the variant that uses the \verb|tslm| function from the \verb|forecast| package, whereas the Dynamic-Harmonic-Regression (A) variant uses the \verb|auto.arima| function from the \verb|forecast| package.

Based on the training paradigms defined in Section~\ref{sec:tp}, we introduce variants of the LSTM-MSNet methodology for our comparative evaluation. These methods are summarised in Table \ref{tab:variants}, represented by the corresponding training paradigm and decomposition technique. Moreover, as the baseline, we use LSTM-MSNet, excluding the seasonal decomposition phase. In other words, we use original observations of the time series, without using DS or SE, to train the LSTM-MSNet. This is referred to as LSTM-Baseline in our experiments.

\begin{table}[!tb]
\caption{Summary of LSTM-MSNet variants}
\centering
\resizebox{\columnwidth}{!}{%
\begin{tabular}{lcl}
	\toprule
	LSTM Variant            			&Training Paradigm 	&Decomposition Technique\\ \hline
	LSTM-MSTL-DS			  			&DS					&MSTL (\textit{s.window}=periodic)\\
	LSTM-MSTL-7-DS						&DS					&MSTL (\textit{s.window}=7)\\
	LSTM-STR-DS							&DS					&STR\\
	LSTM-Prophet-DS						&DS					&Prophet \\
	LSTM-TBATS-DS						&DS					&TBATS \\
	LSTM-Prophet-DS						&DS					&Prophet \\
	LSTM-MSTL-SE						&SE					&MSTL (\textit{s.window}=periodic) \\
	LSTM-MSTL-7-SE						&SE					&MSTL (\textit{s.window}=7)\\
	LSTM-STR-SE							&SE					&STR	\\
	LSTM-Prophet-SE						&SE					&Prophet \\
	LSTM-TBATS-SE						&SE					&TBATS \\
	LSTM-Prophet-SE						&SE					&Prophet \\
	LSTM-Fourier-SE						&SE					&Fourier Transformation \\
	LSTM-Fourier-SE ($k = 1$)			&SE					&Fourier Transformation \\
	\hline
\end{tabular}
}
\label{tab:variants}
\end{table}

\subsection{Computational Performance}

We also report the computational costs in execution time of our proposed LSTM-MSNet variants and the benchmark models on the aggregated hourly energy dataset. The experiments are run on an Intel(R) i7 processor (3.2 GHz), with 2 threads per core, 6 cores per socket, and 64GB of main memory (See Table~\ref{tab:time}).

\subsection{RNN Architecture Comparison}

We perform a set of preliminary experiments to assess the compatibility of our proposed framework with other RNN architectures. To compare against the LSTM cell, we use popular RNN units, namely the Elman RNN  (ERNN) cell \cite{Elman1990-my} and the Gated Recurrent Unit (GRU) cell \cite{Cho2014-or}, which are commonly used for sequence modelling tasks. Table V shows the evaluation summary of these RNN architectures on the benchmark datasets. In our experiments, we use the MSTL-DS and the Fourier-SE ($k = 1$) variants of each RNN architecture to represent DS and SE training paradigms, respectively. For each dataset, the results of the best performing method(s) are marked in boldface.

\begin{table}[!htb]
\caption{RNN Architecture Comparison}
\begin{center}
\resizebox{\columnwidth}{!}{
\begin{tabular}{lcccccccc}
\toprule
& \multicolumn{2}{c}{ \scriptsize{M4}} & \multicolumn{2}{c}{\scriptsize{AusGrid-Energy (Half-hourly)}} & \multicolumn{2}{c}{\scriptsize{AusGrid-Energy (Hourly)}}& \multicolumn{2}{c}{\scriptsize{Traffic}} \\
 & \multicolumn{2}{c}{sMAPE} & \multicolumn{2}{c}{sMAPE} & \multicolumn{2}{c}{sMAPE}& \multicolumn{2}{c}{sMAPE} \\
Method  & Mean & Median & Mean & Median &  Mean & Median & Mean & Median \\
\hline
LSTM-MSTL-DS   &\textbf{0.1069} &\textbf{0.0652} &0.1587 &0.1498 &0.3481 &0.3415 &0.0226 &0.0188\\ 
GRU-MSTL-DS  &0.1116 &0.0723 &0.1635 &0.1558 &0.3420 &0.3371 &0.0224 &0.0188\\ 
ERNN-MSTL-DS   &0.1109 &0.0719 &0.1633 &0.1571 &0.3421 &0.3371 &0.0221 &0.0184\\
LSTM-Fourier-SE ($k = 1$)  &0.1387 &0.0566 &0.1525 &0.1428 &\textbf{0.2590} &\textbf{0.2473} &\textbf{0.0151} &\textbf{0.0116}\\ 
GRU-Fourier-SE ($k = 1$)  &0.1376 &0.0582 &\textbf{0.1513} &\textbf{0.1379} &0.2608 &0.2513 &0.0158 &0.0121\\
ERNN-Fourier-SE ($k = 1$) &0.1381 &0.0583 &\textbf{0.1513} &0.1392 &0.2611 &0.2532 &0.0182 &0.0141\\
\hline
\end{tabular}
}
\label{tab:rnnresults}
\end{center}
\end{table}

We observe that the overall $p$-value of the Friedman rank sum test for M4 dataset is $8.57 \times 10^{-11}$, which is highly significant. The LSTM-MSTL-DS method performs best and achieves significantly better results than the other methods. The overall $p$-value for the AusGrid-Energy (Half-hourly) is $2.89 \times 10^{-6}$. Therefore, GRU-Fourier-SE ($k = 1$) performs significantly better than the rest of the methods. Also, for the AusGrid-Energy (Hourly) dataset, the Friedman rank sum test gives an overall $p$-value of $2.16 \times 10^{-10}$, and LSTM-Fourier-SE ($k = 1$) is used as the control method, which performs best. Finally, the overall $p$-value for the Traffic dataset is $ p < 10^{-10}$. Therefore, the differences among the methods are significant. Also, according to Table~\ref{tab:rnnresults} we see that the LSTM-Fourier-SE ($k = 1$) achieves significantly better results than the other methods. It is also evident that our framework can be employed with any other RNN architecture such as the ERNN cell and the GRU cell. However, according to Table~\ref{tab:rnnresults}, overall we see that the LSTM cell based variants achieve competitive results compared to the ERNN and GRU cells. Therefore, we use the LSTM cell as the primary RNN architecture in our experiments.

\subsection{Results}

Table~\ref{tab:m4results} summarises the evaluation results of all the LSTM-MSNet variants and benchmarks for the 414 hourly series of the M4 competition dataset, ordered by the first column, which is the Mean sMAPE. For each column, the results of the best performing method(s) are marked in boldface. According to Table~\ref{tab:m4results}, the proposed LSTM-MSTL-DS variant obtains the best Mean SMAPE, while FFORMA achieves the best Median SMAPE. We see that regarding the mean MASE, Dynamic-Harmonic-Regression with the \textit{auto.arima} variant performs better than the rest of the benchmarks, whereas FFORMA outperforms the proposed LSTM variants, in terms of the median MASE. Also, on average, the LSTM-MSNet variants with the DS training paradigm achieve better accuracies compared to those of SE. Furthermore, the proposed LSTM-MSTL-DS variant consistently outperforms many state-of-the-art methods, such as TBATS, Prophet, and Dynamic-Harmonic-Regression variants in terms of Mean sMAPE. It is also noteworthy to mention that all the proposed LSTM-MSNet variants outperform our baseline model, LSTM-Baseline, in terms of mean sMAPE and mean MASE.

\begin{table}[!tb]
\caption{M4 dataset results}
\centering
\resizebox{\columnwidth}{!}{%
\begin{tabular}{lcccc}
\hline
	 		Method           				&Mean sMAPE	 		&Median sMAPE  &Mean MASE 	&Median MASE\\ \hline
			LSTM-MSTL-DS					&\bf 0.1069			&0.0652			&0.7131		&0.6340 \\
			LSTM-STR-DS						&0.1099				&0.0734			&0.7528		&0.6554\\
			LSTM-Prophet-DS					&0.1131				&0.0623			&0.8032		&0.6511\\
			FFORMA							&0.1151				&\bf 0.0404		&0.7131		&\bf 0.4750\\
			LSTM-MSTL-7-DS					&0.1187				&0.0795			&0.7840		&0.7092\\
			LSTM-TBATS-DS					&0.1241				&0.0589			&0.8734		&0.7036\\
			Dynamic-Harmonic-Regression (A)	&0.1253				&0.0524			&\bf 0.6937	&0.5732 \\
			LSTM-MSTL-SE					&0.1275				&0.0518			&0.9642		&0.6986\\
			LSTM-STR-SE						&0.1285				&0.0559			&0.9308		&0.6581\\
			TBATS							&0.1309				&0.0477			&0.7781		&0.5390\\
			Prophet							&0.1334				&0.0689			&0.9685		&0.7415	\\		
			LSTM-Fourier-SE					&0.1345				&0.0555			&0.9541		&0.6606\\
			LSTM-TBATS-SE					&0.1368				&0.0568			&0.9675		&0.6492\\
			LSTM-MSTL-7-SE					&0.1388				&0.0631			&0.9796		&0.7334\\
			LSTM-Prophet-SE					&0.1393				&0.0591			&0.9737		&0.6732\\
			LSTM-Fourier-SE ($k = 1$)		&0.1387				&0.0566			&1.0176		&0.6581\\
			LSTM-Baseline					&0.1427				&0.0569			&1.0675		&0.7243\\	
			Dynamic-Harmonic-Regression (T)	&0.1611				&0.1456			&0.7821		&0.6439 \\
			\hline
\end{tabular}
}
\label{tab:m4results}
\end{table}

\begin{table}[!tb]
\caption{Significance testing for M4 dataset}
\centering
\footnotesize
\begin{tabular}{llrrr}
	\toprule
	Method			&$p_{Hoch}$&\\ \hline
	FFORMA			&-\\
	\hline	
	LSTM-MSTL-DS		&8.74 $\times$ $10^{-6}$ \\
	LSTM-STR-DS		&8.74 $\times$ $10^{-6}$ \\	
	Dynamic-Harmonic-Regression (A)				&7.07 $\times$ $10^{-7}$ \\
	TBATS				&2.75 $\times$ $10^{-8}$ \\	
	LSTM-MSTL-7-DS			&1.01 $\times$ $10^{-9}$ \\
	LSTM-Prophet-DS		&3.32 $\times$ $10^{-13}$ \\
	LSTM-STR-SE		&2.41 $\times$ $10^{-21}$ \\
	LSTM-TBATS-SE		&1.23 $\times$ $10^{-22}$ \\
	LSTM-TBATS-DS			&1.68 $\times$ $10^{-25}$ \\
	LSTM-Fourier-SE	&2.80 $\times$ $10^{-30}$ \\
	LSTM-Fourier-SE ($k = 1$)	 &1.75 $\times$ $10^{-35}$ \\
	LSTM-MSTL-SE		&1.09 $\times$ $10^{-40}$ \\
	LSTM-Prophet-SE	&1.53 $\times$ $10^{-51}$ \\
	Prophet				&5.10 $\times$ $10^{-61}$ \\
	LSTM-MSTL-7-SE		&6.38 $\times$ $10^{-65}$ \\
	Dynamic-Harmonic-Regression (T)		&1.64 $\times$ $10^{-66}$ \\	
	LSTM-Baseline			    &1.93 $\times$ $10^{-79}$ \\	
	\hline			
\end{tabular}
\label{tab:m4stat}
\end{table}

Table~\ref{tab:m4stat} shows the results of the  statistical testing evaluation. Adjusted $p$-values calculated from the Friedman test with Hochberg’s post-hoc procedure are presented. A horizontal line is used to separate the methods that perform significantly worse than the best performing method. The overall result of the Friedman rank sum test is a $p$-value of $2.91 \times 10^{-10}$, which is highly significant. The FFORMA method performs best and is used as the control method. Also, according to Table~\ref{tab:m4stat}, we see that the FFORMA achieves significantly better results than the other methods. 

Table~\ref{tab:Ausgrid} shows the evaluation summary for the 300 half-hourly series of the AusGrid-Energy dataset. The \textit{NA} values in the table represent models that could not complete the execution within a time frame of 6 days. It can be seen that the proposed LSTM-MSTL-SE variant outperforms all the benchmarks in terms of the Mean sMAPE and Median sMAPE. Meanwhile, the proposed LSTM-Prophet-DS variant achieves the best accuracy with respect to Mean MASE and Median MASE. Here, in the majority of the cases, the LSTM variants that use SE as the training paradigm obtain better forecasts, which is contrary to our previous findings from the M4 competition dataset. Also, several LSTM-MSNet variants with the DS training paradigm display poor performance compared to the LSTM-Baseline. Most importantly, we observe that the proposed LSTM variants LSTM-MSTL-SE and LSTM-Prophet-DS consistently surpass the current state of the art in all performance metrics.

\begin{table}[!tb]
\caption{AusGrid-Energy (Half-hourly) dataset results}
\centering
\resizebox{\columnwidth}{!}{%
\begin{tabular}{lcccc}
\hline
	 		Method           				&Mean sMAPE	 		&Median sMAPE  	&Mean MASE 	&Median MASE\\ \hline
	 		LSTM-MSTL-SE					&\bf 0.1475			&\bf 0.1369		&0.7461		&0.5900\\
	 		LSTM-Prophet-DS					&0.1478				&0.1397			&\bf 0.7350	&\bf 0.5809\\
	 		LSTM-TBATS-SE					&0.1479				&0.1374			&0.7471		&0.5984\\
	 		LSTM-Prophet-SE					&0.1511				&0.1397			&0.7589		&0.6071\\
	 		LSTM-MSTL-7-SE					&0.1523				&0.1422			&0.7651		&0.6115\\
	 		LSTM-Fourier-SE ($k = 1$)		&0.1525				&0.1428			&0.7694		&0.6052\\
	 		LSTM-Fourier-SE					&0.1527				&0.1414			&0.7676		&0.6130\\
	 		LSTM-Baseline					&0.1561				&0.1430			&0.7838		&0.6263\\
	 		LSTM-MSTL-DS					&0.1587				&0.1498		   	&0.7729		&0.6458 \\
	 		TBATS							&0.1597				&0.1467			&0.8221		&0.6506\\
	 		LSTM-MSTL-7-DS					&0.1620				&0.1538			&0.7932		&0.6599\\
	 		FFORMA							&0.1808				&0.1708			&0.9615		&0.6953\\
	 		Prophet							&0.1848				&0.1766			&0.8935		&0.7346\\
	 		Dynamic-Harmonic-Regression	(A)	&0.1919				&0.1808			&0.9138		&0.7599 \\
	 		Dynamic-Harmonic-Regression	(T)	&0.2059				&0.1847			&0.9773		&0.8567 \\
	 		LSTM-TBATS-DS					&0.3092				&0.3014			&1.3011		&1.0601\\
			LSTM-STR-DS						&NA					&NA				&NA			&NA	\\
			LSTM-STR-SE						&NA					&NA				&NA			&NA	\\
			\hline
\end{tabular}
}
\label{tab:Ausgrid}
\end{table}

\begin{table}[!tb]

\caption{Significance testing for AusGrid-Energy (Half-hourly) dataset}
\centering
\footnotesize
\begin{tabular}{llrrr}
	\toprule
	Method				&$p_{Hoch}$&\\ \hline
	LSTM-MSTL-SE 		&-\\
	LSTM-TBATS-SE		&0.600 \\
	LSTM-Prophet-DS		&0.233 \\
	\hline	
	LSTM-Prophet-SE		&8.89 $\times$ $10^{-4}$ \\	
	LSTM-MSTL-7-SE		&7.62 $\times$ $10^{-6}$ \\
	LSTM-Fourier-SE		&6.69 $\times$ $10^{-7}$ \\
	LSTM-Fourier-SE ($k = 1$)	&4.56 $\times$ $10^{-7}$ \\
	TBATS					&7.48 $\times$ $10^{-8}$ \\	
	LSTM-MSTL-DS			&2.76 $\times$ $10^{-9}$ \\
	LSTM-Baseline-DS			&5.35 $\times$ $10^{-15}$ \\
	LSTM-MSTL-7-DS				&1.19 $\times$ $10^{-22}$ \\
	FFORMA					&5.96 $\times$ $10^{-55}$ \\
	Prophet					&4.65 $\times$ $10^{-62}$ \\
	Dynamic-Harmonic-Regression (A)	    &5.54 $\times$ $10^{-64}$ \\
	Dynamic-Harmonic-Regression (T)		&3.32 $\times$ $10^{-64}$ \\
	LSTM-TBATS-DS			&1.94 $\times$ $10^{-91}$ \\
	\hline			
\end{tabular}
\label{tab:solarhalfhourlystat}
\end{table}

The Friedman rank sum test gives an overall $p$-value of $ p < 10^{-10}$. Therefore, the differences among the benchmarks are highly significant. According to Table~\ref{tab:solarhalfhourlystat}, we see that the LSTM-MSTL-SE performs best and is used as the control method. Moreover, the LSTM-MSNet variants, LSTM-TBATS-SE, LSTM-Prophet-DS do not perform significantly worse than the control method. 

Table~\ref{tab:Ausgrid1} provides the results for the evaluations on the aggregated hourly time series of the AusGrid-Energy dataset. We see that the proposed LSTM-Fourier-SE ($k = 1$) variant achieves the best results on each performance metric, and outperforms the rest of the benchmarks. Also, among the proposed variants, we observe that the LSTM-MSNet variants with the SE training paradigm outperform their counterparts, the LSTM-MSNet variants with the DS training paradigm. Furthermore, the LSTM-Baseline method performs better than the LSTM-MSNet variants with DS training paradigm. Nevertheless, consistent with our previous findings from Table~\ref{tab:Ausgrid}, the proposed LSTM-MSNet variants outperform the current state-of-the-art techniques; FFORMA, TBATS, and Prophet.

\begin{table}[!tb]
\caption{AusGrid-Energy (Hourly) dataset results}
\centering
\resizebox{\columnwidth}{!}{%
\begin{tabular}{lcccc}
\hline
	 		Method           				&Mean sMAPE	 		&Median sMAPE  	&Mean MASE 	&Median MASE\\ \hline
	 		LSTM-Fourier-SE ($k = 1$)		&\bf 0.2590			&\bf 0.2473	 	&\bf 0.7189 &\bf 0.6455\\
			LSTM-MSTL-SE					&0.2638				&0.2626		 	&0.7286	 	&0.6652\\
			LSTM-Prophet-SE					&0.2653				&0.2575		 	&0.7332	 	&0.6819\\
			LSTM-TBATS-SE					&0.2665				&0.2676		 	&0.7346	 	&0.6753\\
			LSTM-Fourier-SE					&0.2672				&0.2572		 	&0.7399	 	&0.6790\\
			LSTM-Baseline					&0.2685				&0.2660			&0.7408		&0.6852\\
			FFORMA							&0.2692				&0.2613		 	&0.7360	 	&0.6573\\
			LSTM-Prophet-DS					&0.2749				&0.2761		 	&0.7526	 	&0.6913\\
			LSTM-MSTL-7-SE					&0.2884				&0.2785		 	&0.7930	 	&0.7219\\
			LSTM-MSTL-7-DS					&0.3172				&0.3059		 	&0.8250	 	&0.7790\\
			TBATS							&0.3177				&0.3072		 	&0.8179	 	&0.7861\\
			Prophet							&0.3390				&0.3369	     	&0.8757	 	&0.8156\\
			LSTM-TBATS-DS					&0.3414				&0.2971		 	&1.0640	 	&0.7458\\
			LSTM-MSTL-DS					&0.3480				&0.3415		 	&0.9123	 	&0.8550 \\
			Dynamic-Harmonic-Regression	(T)	&0.3530				&0.3460			&0.9015		&0.8692 \\
			Dynamic-Harmonic-Regression (A)	&NA					&NA				&NA			&NA \\
			LSTM-STR-DS						&NA					&NA				&NA			&NA	\\
			LSTM-STR-SE						&NA					&NA				&NA			&NA	\\
			\hline
\end{tabular}
}
\label{tab:Ausgrid1}
\end{table}

\begin{table}[!tb]
\caption{Significance testing for AusGrid-Energy (Hourly) dataset}
\centering
\footnotesize
\begin{tabular}{llrrr}
	\toprule
	Method					&$p_{Hoch}$&\\ \hline
	LSTM-Fourier-SE ($k = 1$)	&- \\
	LSTM-MSTL-SE 		&0.265\\
	LSTM-TBATS-SE		&0.128 \\
	LSTM-Prophet-SE	&0.128 \\
	LSTM-Fourier-SE	&0.128\\
	FFORMA				&0.084\\
	\hline	
	LSTM-Prophet-DS		&0.041 \\
	LSTM-Baseline		&0.016 \\
	LSTM-TBATS-DS			&2.70 $\times$ $10^{-9}$ \\
	LSTM-MSTL-7-SE		&2.99 $\times$ $10^{-15}$ \\
	TBATS				&3.27 $\times$ $10^{-23}$ \\	
	LSTM-MSTL-7-DS			&2.25 $\times$ $10^{-23}$ \\
	Prophet				&1.98 $\times$ $10^{-44}$ \\
	Dynamic-Harmonic-Regression (T)		&3.46 $\times$ $10^{-45}$ \\
	LSTM-MSTL-DS				&1.40 $\times$ $10^{-57}$ \\
	\hline			
\end{tabular}
\label{tab:solarhourlystat}
\end{table}

The overall result of the Friedman rank sum test is a $p$-value of $2.76 \times 10^{-10}$, which means the results are highly significant. According to Table~\ref{tab:solarhourlystat}, the LSTM-Fourier-SE ($k = 1$) performs best and is used as the control method. We see that the LSTM-MSNet variants, LSTM-MSTL-SE, LSTM-TBATS-SE, LSTM-Prophet-SE, LSTM-Fourier-SE and FFORMA do not perform significantly worse than the control method.

\begin{table}[!tb]
\caption{Traffic dataset results}
\centering
\resizebox{\columnwidth}{!}{%
\begin{tabular}{lcccc}
\hline
	 		Method           				&Mean sMAPE	 		&Median sMAPE  	&Mean MASE 	&Median MASE\\ \hline
	 		LSTM-Fourier-SE ($k = 1$)		&\bf 0.0151			&\bf 0.0116	 	&\bf 0.6102 &\bf 0.5035\\
	 		LSTM-Fourier-SE					&0.0179				&0.0143		 	&0.7213	 	&0.6178\\
			LSTM-MSTL-SE					&0.0177				&0.0142		 	&0.7107	 	&0.6159\\
			LSTM-TBATS-SE					&0.0186				&0.0151		 	&0.7467	 	&0.6495\\
			LSTM-Prophet-SE					&0.0194				&0.0157		 	&0.7828	 	&0.6744\\
			LSTM-Baseline					&0.0193				&0.0158			&0.7853		&0.6777\\
			LSTM-MSTL-7-SE					&0.0213				&0.0179		 	&0.8760	 	&0.7801\\
			LSTM-Prophet-DS					&0.0219				&0.0191		 	&0.8951	 	&0.7928\\
			LSTM-MSTL-DS					&0.0232				&0.0196		 	&0.9529	 	&0.8321 \\
			FFORMA							&0.0245				&0.0190		 	&0.9177	 	&0.7582\\
			LSTM-STR-DS						&0.0249				&0.0209			&0.9918		&0.8710	\\
			LSTM-MSTL-7-DS					&0.0303				&0.0273		 	&1.2619	 	&1.1576\\
			TBATS							&0.0247				&0.0215		 	&1.0412	 	&0.8861\\
			LSTM-STR-SE						&0.0251				&0.0217			&1.0377		&0.9334	\\
			Dynamic-Harmonic-Regression (A)	&0.0303				&0.0259			&1.1987		&1.0555 \\
			Prophet							&0.0332				&0.0298	     	&1.3248	 	&1.2061\\
			Dynamic-Harmonic-Regression	(T)	&0.0335				&0.0286			&1.1570		&1.1812 \\
			LSTM-TBATS-DS					&0.0789				&0.0773		 	&3.1593	 	&2.9352\\
			\hline
\end{tabular}
}
\label{tab:traffic}
\end{table}

\begin{table}[!tb]
\caption{Significance testing for Traffic dataset}
\centering
\footnotesize
\begin{tabular}{llrrr}
	\toprule
	Method						&$p_{Hoch}$&\\ \hline
	LSTM-Fourier-SE ($k = 1$)	&- \\
	\hline	
	LSTM-MSTL-SE 				&1.44 $\times$ $10^{-15}$\\
	LSTM-Fourier-SE				&4.32 $\times$ $10^{-21}$\\
	LSTM-TBATS-SE				&6.06 $\times$ $10^{-42}$\\
	LSTM-Baseline				&6.79 $\times$ $10^{-71}$\\
	LSTM-Prophet-SE				&8.77 $\times$ $10^{-84}$ \\
	LSTM-MSTL-7-SE				&$p < 10^{-100}$ \\
	FFORMA						&$p < 10^{-100}$ \\
	LSTM-MSTL-DS				&$p < 10^{-100}$ \\
	LSTM-Prophet-DS				&$p < 10^{-100}$ \\
	TBATS						&$p < 10^{-100}$ \\
	LSTM-STR-DS					&$p < 10^{-100}$ \\
	LSTM-STR-SE					&$p < 10^{-100}$ \\
	Dynamic-Harmonic-Regression (A) &$p < 10^{-100}$ \\
	Dynamic-Harmonic-Regression (T)	&$p < 10^{-100}$ \\
	LSTM-MSTL-7-DS					&$p < 10^{-100}$ \\
	Prophet							&$p < 10^{-100}$ \\
	LSTM-TBATS-DS					&$p < 10^{-100}$ \\
	\hline			
\end{tabular}
\label{tab:traffichourlystat}
\end{table}

Table~\ref{tab:traffic} provides the results for the evaluations on the hourly time series of the San Francisco traffic dataset. We observe that the results are similar to the previous findings from Table~\ref{tab:Ausgrid1}, where the proposed LSTM-Fourier-SE ($k = 1$) variant outperforms the rest of the benchmarks. Moreover, the LSTM-MSNet variants with the SE training paradigm achieves better results, compared to the LSTM-MSNet variants with the DS training paradigm. Also, consistent with our observations from the AusGrid-Energy datasets, the LSTM-Baseline method performs better than the LSTM-MSNet variants with DS training paradigm. Furthermore, the majority of the LSTM-MSNet variants outperform the statistical benchmarks, FFORMA, TBATS, and Prophet.

The Friedman rank sum test gives an overall $p$-value of $ p < 10^{-10}$, which means the results are highly significant. According to Table XIII, the LSTM-Fourier-SE ($k = 1$) performs best and is used as the control method. Also, we see that the LSTM-Fourier-SE ($k = 1$) achieves significantly better results than the other methods. 

Also, with respect to computational cost of the proposed variants and benchmarks, according to Table~\ref{tab:time}, except for LSTM-TBATS-DS and LSTM-TBATS-SE, which use TBATS as the decomposition technique, we see that the proposed LSTM-MSNet variants have a lower execution time compared to TBATS and FFORMA. The Dynamic-Harmonic-Regressio\-n\- (T) variant and Prophet are more computationally efficient than the LSTM-MSNet variants.
However, we notice that both these methods do not display competitive results compared to the LSTM-MSNet variants, according to Tables~\ref{tab:m4results}, \ref{tab:Ausgrid}, and \ref{tab:Ausgrid1}. In contrary, LSTM-MSNet promises to deliver better performance than TBATS and FFORMA, with respect to both accuracy and computation time.

\begin{table}[tb!]
\caption{Computational Summary - AusGrid-Energy Dataset: Hourly (in Minutes)}
\centering
\resizebox{\columnwidth}{!}{%
\begin{tabular}{lccc}
\hline
Method &Pre-processing &Model-Training \& Post-processing &Total-time\\
\hline
Dynamic-Harmonic-Regression	(T) &- &4 &4\\
Prophet &- &90 &90\\
LSTM-Baseline &10 &140 &150\\
LSTM-MSTL-DS &88 &120 &208\\
LSTM-MSTL-7-DS &88 &120 &208\\
LSTM-Fourier-SE ($k = 1$) &30 &180 &210\\
LSTM-Fourier-SE &34 &180 &214\\
LSTM-MSTL-SE &74 &180 &254\\
LSTM-MSTL-7-SE &74 &180 &254\\
LSTM-Prophet-DS &210 &120 &330\\
LSTM-Prophet-SE &200 &180 &380\\
TBATS &- &2304 &2304\\
LSTM-TBATS-DS &2332 &120 &2452\\
LSTM-TBATS-SE &2318 &180 &2498\\
FFORMA &- &4320 &4320\\
\hline
\end{tabular}
}
\label{tab:time}
\end{table}

\subsection{Discussion}

It can be seen that the LSTM-Baseline results for the M4 and real-world datasets are contradictory. Even though the LSTM-Baseline model cannot outperform both DS and SE learning paradigms of LSTM-MSNet in the M4 dataset, it obtains better results than the DS learning paradigm in AusGrid-Energy and Traffic datasets. These results can be interpreted by the seasonal characteristics present in these datasets (Fig.~\ref{seasonalpatterns}). As discussed in Section~\ref{sec:datasets}, the seasonal components of the M4 dataset are less homogeneous. So, in this scenario, learning the seasonality directly from the original time series becomes difficult for LSTM-MSNet. Hence, removing the seasonal components prior to training (DS learning paradigm) has positively contributed towards the LSTM-MSNet results in the M4 dataset. This also explains the poor results of the LSTM-Baseline variant, which attempts to learn seasonality directly from the time series, without any assistance of DS and SE. We also note that lengths of the series are likely to have an effect here, in the sense that for longer series, the amount of data is sufficient to learn seasonal patterns directly, whereas for shorter series, deseasonalisation is beneficial. However, we do not analyse this in our current study.
Furthermore, we observe the highly competitive results of FFORMA in the M4 dataset. FFORMA was the second-best performing method in the overall M4 competition, so it can be arguably seen as optimised for this particular dataset. Also, due to its nature as being an ensemble of simpler univariate techniques, it is inherently more suitable for this situation of a dataset of inhomogeneous series, whereas on the AusGrid-Energy and Traffic datasets, it is less performant. There, due to the high presence of homogeneous seasonality, exclusive learning of the seasonality becomes viable for LSTM-MSNet. As a result, LSTM-Baseline achieves better results compared to LSTM-MSNet with the DS learning paradigm. However, in this scenario, we observe that the variants of LSTM-MSNet with the SE learning paradigm, achieve better results than the LSTM-Baseline model. This suggests that the extracted seasonal components have supplemented the LSTM-MSNet training procedure, in the form of external variables that assist to determine the trajectory of the multiple seasonal cycles.

In terms of decomposition techniques, MSTL, STR, and Prophet are the best performing methods in the M4 dataset. Whereas, on the AusGrid-Energy and Traffic datasets, we see that the LSTM-MSNet variants with MSTL and Prophet decomposition techniques give better results. Also, the STR based LSTM-MSNet variants are unstable on the AusGrid-Energy datasets, which can be attributed to the longer lengths of the time series. Furthermore, according to Table~\ref{tab:time}, the MSTL and Prophet methods are computationally efficient decomposition techniques compared to TBATS.

These results indicate that our proposed LSTM-MSNet forecasting framework can be easily adapted, depending on the seasonal characteristics in a group of time series. The LSTM-MSNet with the DS learning paradigm is more suitable for situations where the origins of the time series are unknown and time series are inhomogeneous. Whereas its counter part, SE is better if the time series are homogeneous and share similar shapes of the seasonal components. Another important finding from this study is that the RNNs are competitive in situations where groups of time series are highly homogeneous. As discussed in Section~\ref{sec:intro}, this is the case in many real-world applications. However, in situations like in the M4 dataset, where the groups of time series exhibit highly heterogeneous characteristics, better competitiveness of the RNNs can be achieved by applying additional preprocessing steps such as deseasonalisation (DS learning paradigm) that supplements the RNN training procedure. Apart from these observations, we also see that for longer time series, the majority of the LSTM-MSNet variants are computationally more efficient than the state-of-the-art univariate forecasting techniques, such as TBATS and FFORMA.

\section{Conclusion}
\label{sec:conclusion}

In this paper, we have presented the LSTM-MSNet methodology, a novel, three-layered forecasting framework that is capable of forecasting a group of related time series with multiple seasonal cycles. Our methodology is based on time series decomposition and LSTM recurrent neural networks, to overcome the limitations of the current univariate state-of-the-art models by training a unified model that exploits key structures, behaviours, and patterns common within a group of time series.

We have utilised a series of decomposition techniques from the literature to extract the various forms of seasonal components in time series. Moreover, we have also discussed two training paradigms, the Deseasonalised approach and the Seasonal Exogenous approach, highlighting how these decomposition techniques can be used to supplement the LSTM learning procedure. We have identified that the choice of these learning paradigms can be determined by the characteristics of the time series, where a deseasonalised approach is more suitable for situations where the series have different seasonal patterns, and the start/end dates of the series are different. Whereas its counter part, the seasonal exogenous approach, is better if the time series are homogeneous and share similar shapes of the seasonal components. In general, MSTL, and Prophet are stable decomposition techniques for both shorter and longer time series, and achieve competitive results with a lower computational cost.
We have evaluated the proposed forecasting framework using a competition dataset, two energy consumption time series datasets, and Traffic dataset that contain multiple seasonal patterns. Also, with respect to accuracy and computational time, we observe that our framework can be a competitive approach among the current state-of-the-art in his research space. Furthermore, somewhat contrary to widespread beliefs, the globally trained LSTM-MSNet can be computationally more efficient than many univariate forecasting methods.

As a possible future work, a hybrid version of this approach can be introduced to handle seasonalities in longer time series. Here, the deseasonalised approach can be used to model shorter seasonalities, whereas the seasonal exogenous approach can be applied to address the longer seasonalities. 

\section{Acknowledgements}
\label{sec:ach}

This research was supported by the Australian Research Council under grant DE190100045, by a Facebook Statistics for Improving Insights and Decisions research award, by Monash Institute of Medical Engineering seed funding, and by the MASSIVE - High performance computing facility, Australia.

\bibliographystyle{IEEEtran}
\bibliography{reference}

\end{document}